%%%%%%%%%%%%%%%%%%%%%%%%%%%%%%%%%%%%%%%%%%%%%%%%%%%%%%%%%%%%%%%
%%  CLAN STRUCTURE ANALYSIS AND RAPIDITY GAP PROBABILITY     %%
%%                                              DFTT 28/94   %%
%%  by S. Lupia, A. Giovannini and R. Ugoccioni              %%
%%  Dipartimento di Fisica Teorica - Universita' di Torino   %%
%%  and I.N.F.N. Sezione di Torino                           %%
%%  via P. Giuria 1 -  10125 Torino - Italy                  %%
%%  e-mails: lupia@to.infn.it   giovannini@to.infn.it        %%
%%           ugoccioni@to.infn.it                            %%
%%%%%%%%%%%%%%%%%%%%%%%%%%%%%%%%%%%%%%%%%%%%%%%%%%%%%%%%%%%%%%%
\magnification=1200
\tolerance=500
\hsize 15.0truecm\hoffset 0.7truecm      % size for camera ready papers
\vsize 666truept          
\baselineskip=15pt                       % 1 1/2 line spacing
\parindent 0.9truecm
\outer\def\beginsection#1\par{\medbreak\bigskip
      \message{#1}\leftline{\bf#1}\nobreak\medskip\vskip-\parskip
      \indent}

\def\ref#1{[#1]}             % use pseudo-references at first, then
                             % change to numbers.

\ifx\usepostscript\undefinedcs  %here \usepostscript is not defined
\font\caps=cmcsc10              % small caps font
\else
\font\caps=ptmrc at 10pt        % small postscript caps font
\fi
               % calligraphic P(Q_0,Q_1|Q)
          % integr.symbol with limits on top&bottom
\def\ee{$e^+e^-$}               % e+ e- (annihilations...)
             % GeV is roman inside mathmode

               % NB parameters: must be used in math mode
               % only, or it won't work!

  % <n>: requires math mode.

\def\delphi{{\caps Delphi}}

        %math mode only
\def\NF{{\cal N}_{\kern -1.9pt f}}     %math mode only
\def\NC{{\cal N}_{\kern -1.7pt c}}     %math mode only
       %math mode only

\def\roots#1{$\sqrt{s} = #1$ GeV}       % the well abused root s = xx GeV...
\def\pt{{p\kern -.2pt\lower 4pt\hbox{\fivei T}}}    %works well enough
                                                    %needs math mode
          % math mode only
          % math mode only

\def\pl{{p\kern -.2pt\lower 4pt\hbox{\fivei L}}}
          % requires math mode
\def\Dy{\Delta y}

%define less than or approx. / greater than or approx. - math mode
\def\nostrocostrutto#1\over#2{\mathrel{\mathop{\kern 0pt \rlap 
  {\raise.2ex\hbox{$#1$}}}
  \lower.9ex\hbox{\kern-.190em $#2$}}}
   %less or around ...
   %greater or around...

\def\TEVATRON{{\caps Tevatron}}
\def\HERA{{\caps Hera}} 
\def\Ngc{\bar N_{\hbox{\sevenrm g-clan}}}
\def\ngc{\bar n_{\hbox{\sevenrm c,g-clan}}}
\outer\def\subsection#1\par{\medbreak\bigskip
      \message{#1}\leftline{\it#1}\nobreak\medskip\vskip-\parskip
      \indent}
\parindent 0.9truecm
\pageno=0
\begingroup 
\nopagenumbers
\null
\vskip -1ex
\baselineskip=14pt
\rightline{DFTT 28/94}
\rightline{July 1994}

\vskip 4.truecm
\centerline{\bf CLAN STRUCTURE ANALYSIS AND RAPIDITY GAP PROBABILITY}
\vskip 1truecm   %(1.5truecm in camera ready)
\centerline {S.\ Lupia, \quad A.\ Giovannini, \quad R.\ Ugoccioni}
\smallskip
\centerline {\it Dipartimento di Fisica Teorica, Universit\`a di Torino}
\centerline {\it and INFN, Sezione di Torino}
\centerline {\it via P. Giuria 1, 10125 Torino, Italy}
\vskip 1.4truecm  %(2true)

\parindent 0cm
\footnote{}{E-mail addresses: giovannini@to.infn.it, lupia@to.infn.it,
ugoccioni@to.infn.it}
\footnote{}{\it Work supported in part by M.U.R.S.T.\ 
(Italy) under Grant 1993}
\parindent 1cm

\vfil
\midinsert
\parindent 1.5truecm
\narrower
\centerline{ {\bf ABSTRACT} }
% TEXT OF ABSTRACT GOES HERE.
\smallskip
\noindent 
Clan structure analysis in rapidity intervals is generalized from 
negative binomial multiplicity distribution to the wide class of 
compound Poisson distributions. The link of generalized clan structure 
analysis with correlation functions is also established. 
These theoretical results are then applied to minimum bias events and 
evidentiate new interesting features, which can be inspiring and useful 
in order to discuss data on rapidity gap probability at \TEVATRON\ and 
\HERA.

\endinsert

\vfill\eject
\endgroup

\beginsection Introduction 

Clan structure analysis puzzled experts since its introduction in 
multiparticle dynamics. The real question is: is clan structure analysis 
simply a new parametrization or, in view of the regularities which it 
reveals in different classes of reactions, it has a deeper physical 
insight? In \ref{1} it has been shown that the inverse of the 
average number of particles per clan is exactly the {\it void scaling 
function}, which was introduced in order to test hierarchical structure 
of correlations. The behavior of the average number of particles per 
clan provides therefore information on the structure of 
correlations functions in multiparticle 
production. In this paper, starting from some ideas developed in 
\ref{1}, we show that also the average number of clans 
in a region of phase space has an 
suggestive physical meaning: it is simply linked to the probability to 
detect no particles in that region, as well as to the 
normalized factorial cumulants generating function in the same region. 
The average number of clans provides therefore information both on 
rapidity gap probabilities and on the general features of correlation 
functions. It is interesting to remark that these properties are not 
linked to the distribution which motivated clan structure analysis, {\it 
i.e.}, Negative Binomial (NB) Multiplicity Distribution (MD), but are 
common to the whole class of Compound Poisson Distributions (CPD's) (or 
discrete infinitely divisible distributions) to which NB MD belongs.  

In Section 1 we generalize clan structure analysis to CPD's 
and discuss the theorems which establish the above mentioned connections. 
In Section 2 we apply these theorems to the domain of validity of NB 
regularity. Interesting new features are revealed and in particular the 
energy independence of the rapidity gap probability 
in different classes of reactions as well as its leveling in 
large rapidity intervals. 

\beginsection I. Generalized clan structure analysis, correlations and 
rapidity gap probability 

A CPD is fully determined by its generating function, $f_{CPD}(z)$, and 
is described in general by the following equation 
$$
f_{CPD}(z) = e^{\Ngc [g(z)-1]} 
\eqno(1)
$$
where $\Ngc$ is the average number of independent 
intermediate objects generated according to a Poisson distribution; they 
have been called generalized clans (g-clans) in \ref{1} whereas 
$g(z)$ is the particle  generating function for an average g-
clan. Notice that a physical process described by a CPD is a typical two 
steps process: intermediate independent g-clans produced in the first 
step decay into final charged 
particles in the second step following the MD 
corresponding to the generating function $g(z)$; g-clans are indeed groups 
of particles of common origin and each g-clan contains at 
least one particle, {\it i.e.}, according to this definition, particles  
MD of an average g-clan has to be truncated:
$$
g(z) \big|_{z=0} = 0 
\eqno(2)
$$
This description fits the full production process, and therefore
applies directly to full phase space analysis. When examining a rapidity
interval $\Dy$ one should understand that we are not defining a new process
but we are again dealing with the g-clans and particles  described
above. Thus one is lead again to eq.~(1) but now $g(z)$ becomes $g(z;\Dy)$,
{\it i.e.}, it is the generating function of the MD of one g-clan with respect to the
interval $\Dy$, which is in general different for different intervals:
$$
  f(z;\Dy) = e^{\Ngc\, [g(z;\Dy)-1]} \qquad , \qquad
  g(z;\Dy)\big|_{z=0} \equiv q_0(\Dy) \ne 0
  \eqno(3)
$$
In fact $q_0(\Dy)$ is the probability that a g-clan does not produce any particle
within the interval $\Dy$, which is indeed not zero. It is easily seen
that by defining a new generating function $\tilde g(z;\Dy)$:
$$
  \tilde g(z;\Dy) = { g(z;\Dy)-q_0(\Dy) \over 1-q_0(\Dy)}
  \eqno(4)
$$
one can write eq.~(3) as
$$
    f(z;\Dy) = e^{\Ngc(\Dy)\,[\tilde g(z;\Dy)-1]} \qquad , \qquad
  \tilde g(z;\Dy)\big|_{z=0} = 0
  \eqno(5)
$$
Since now $\tilde g(z;\Dy)|_{z=0} = 0$, eq.~(5) involves only those g-clans
which produce at least one particle in the interval $\Dy$. (It should be
pointed out that the form of the distribution used in the standard clan
analysis of data is indeed eq.~(5).)

\noindent
Finally note that eq.s~(3) and (5) are linked by binomial convolution
at {\it g-clan\/ level}, since one finds
$$
    \Ngc(\Dy) = \Ngc\, [1-q_0(\Dy)]
    \eqno(6)
$$
which, in terms of probabilities for the observation of $N'$ g-clans in the
interval $\Dy$, $p_{N'}(\Dy)$, and of $N$ g-clans in full phase space, $p_N$,
corresponds to
$$
    p_{N'}(\Dy) = \sum_{N=N'}^{\infty} {N \choose N'} q_0(\Dy)^{N-N'}
      [1-q_0(\Dy)]^{N'} p_N
    \eqno(7)
$$
Eq.~(6) should be contrasted with the application of binomial convolution
at particle level, which leads\ref{2}, in the case of NBD, to
$$
    \bar n(\Dy) = \bar n\, [1-P_0(\Dy)] \qquad\qquad
      k(\Dy) = k
$$
in sharp contrast to experimental data; on the contrary, 
eq.s~(6) and (7) don't give any
relationship between parameters at particle level, since they don't imply
any relationship between $g(z)$ and $g(z;\Dy)$.

It is well known that a NB MD can be obtained by requiring  
a logarithmic distribution of particles inside an average g-clan.  
NB MD is an appreciated two-parameter MD in multiparticle dynamics; its 
standard parameters are the average number of charged particle, 
$\bar n$, and the parameter $k^{-1}$, which is linked to the dispersion 
$D^2 = \bar {n^2}  - {\bar n}^2$ by the relation $k^{-1} = D^2/{\bar n}^2
 - 1/\bar n = \kappa_2$ ($\kappa_2$ is the second-order normalized 
factorial cumulant). NB distribution has been proposed 
since 1972\ref{3} 
with success for describing experimental data on final charged particle 
MD's in full phase space and later on\ref{4} 
in symmetric rapidity intervals 
in hadron-hadron collisions, and then extended to 
all classes of high energy reactions 
(deep inelastic scattering, \ee\ annihilation and $AA$ collisions)
\ref{5}. 
The interest on the class of CPD's, to which NB belongs, seems 
therefore fully justified. 
It should be added that the interpretation of NB 
regularity\ref{6} led to 
analyze experimental data in terms of the average number of clans, $\bar 
N$ (the name clan was introduced in this framework) and of the average 
number of particles per clan $\bar n_c$ (clan structure analysis). 
The new parametrization is 
linked to the old one in terms of $\bar n$ and $k$ by the following 
equations:
$$
\bar N = k \log \left( 1 +{ \bar n \over k} \right) 
\eqno(8)
$$
$$
\bar n_c = {\bar n \over \bar N} 
\eqno(9)
$$
The interest here is on the connection of the generalized clan 
properties with the probability to detect no particles in a rapidity 
interval $\Dy$ for a generic CPD; the general theory will be 
applied to NB MD in the next section.  
Being in eq.s~(1) and (5) 
the generating function $g(z)$ not a priori specified, 
clan concept results to be much more general than the standard one 
defined by eq.s~(8) and (9). 
Depending on the choice of $g(z)$ one has in 
fact different CPD generating functions $f_{CPD}(z)$. 
Table 1 shows, in addition to NB MD,  
two other CPD's frequently discussed in the literature\ref{7}, 
{\it i.e.},  the composition of a Poisson with a truncated Poisson 
distribution (Thomas distribution\ref{8}) and the composition of a 
Poisson with a truncated geometric distribution 
(P\' olya-Aeppli\ref{9}).  
All these are 
two-parameter distributions. Notice that they 
can be obtained as limiting forms of the composition of a Poisson with a 
 NB distribution, {\it i.e.}, a  three parameters distribution. 
A fourth distribution is shown in Table 1 as a typical example of a 
three parameter distribution,  the Partially Coherent Laser 
Distribution (PCLD)\ref{10}: its generating function 
is the product of the generating functions of NB and P\' olya-Aeppli 
distributions; accordingly, the PCLD belongs to the class of CPD's, a 
fact which has been overlooked in the literature. 

It should be noticed that the belonging to the class of CPD's for a MD can 
be tested either by the sign of the corresponding 
combinants\footnote{*}{Combinants $C_n$ are defined
by $f(z) = \exp [\sum_{n=0}^{\infty} C_n (z^n - 1)]$ such that for 
CPD's one has 
$\bar N = - \log P_0  = \sum_{n=0}^{\infty} C_n$ and $C_n/\bar N$ 
is the probability to have $n$ particles inside an average g-clan; 
therefore one has a CPD {\it iff} $P_0 >0$ and $C_n \ge 0, 
\forall n$.}\ref{11}  or, for two-parameter 
distributions only, 
by the validity of Linked-Pair Ansatz (LPA) for the 
corresponding $n$-particle correlations functions\ref{12} (its 
violation for two-parameters MD's implies that the MD is not a CPD). 

The study of $\Ngc(\Dy)$ and $P_0(\Dy)$ for the 
class of CPD's is based on the following simple theorems. 

Theorem 1.\  
Being by definition 
$$
P_0(\Dy) \equiv f(z;\Dy)\big|_{z=0} 
\eqno(10)
$$
from eq.~(5) it follows that 
$$
P_0(\Dy) = e^{- \Ngc(\Dy)}
\eqno(11)
$$
Eq.~(11) says that for CPD's  
the average number of g-clans in the rapidity interval $\Dy$ 
determines the probability of detecting 
no particles in the same interval.

Theorem 2.\ 
From the expression of the generating function in terms of normalized 
factorial cumulants\ref{7}
$$
f(z;\Dy) = 
\exp \left\{ \sum_{n=1}^{\infty} {\kappa_n(\Dy) \over n!} 
[ \bar n(\Dy) (z-1)]^n \right\}
\eqno(12)
$$
a second powerful theorem on $P_0(\Dy)$  can be proved via 
eq.~(10).
It establishes the link of $P_0(\Dy)$ with the corresponding 
 normalized factorial cumulants in 
the interval $\Dy$, $\kappa_n(\Dy)$. One has in fact
$$
P_0(\Dy) = \exp \left[ \sum_{n=1}^{\infty} 
{[- \bar n(\Dy)]^n \over n!} \kappa_n(\Dy) \right] 
\eqno(13)
$$
where normalized factorial cumulants $\kappa_n(\Dy)$ are $n$-fold integrals of 
the normalized correlation functions $c_n(y_1,\dots,y_n)$:
$$
\kappa_n(\Dy) = \int_{\Dy} dy_1 \dots 
\int_{\Dy} dy_n c_n(y_1,\dots,y_n)
\eqno(14) 
$$
$P_0(\Dy)$ turns out to be determined according to 
eq.s~(13) and (14) by the sum of all $n$-order normalized factorial 
cumulants or,  
equivalently, by the integrals over the interval 
$\Dy$ of corresponding $n$-particle correlation functions. 
Notice that the exponent in eq.~(13) is the normalized factorial cumulant generating 
function. 

Theorem 3.\  Finally it 
can be proved just by inspection of eq.s~(11) and (13) that 
$$
\Ngc(\Dy) = - \sum_{n=1}^{\infty} 
{[- \bar n(\Dy)]^n \over n!} \kappa_n(\Dy) 
\eqno(15)
$$
{\it i.e.}, the average number of g-clans for a CPD in a given rapidity 
interval $\Dy$ can be obtained by calculating the normalized factorial 
cumulants 
generating function in the same interval and vice versa the normalized 
factorial 
cumulant generating function is fully determined by the average number 
of g-clans.

The generality of the above mentioned theorems leads to striking results 
when applied to the class of hierarchical models\ref{13}, {\it i.e.}, to the 
class of models in which $c_n(y_1,\dots,y_n)$ can be expressed as the 
product of $(n-1)$ two-particle correlation functions or, in
terms of normalized factorial cumulants:
$$
\kappa_n(\Dy) = A_n [\kappa_2(\Dy)]^{n-1} 
\eqno(16)
$$
$A_n$ in eq.~(16) is independent of the energy and rapidity interval 
considered, but it depends on the different choices of the generating 
function $g(z;\Dy)$ in eq.~(5). 
Distributions NB, Thomas and P\' olya-Aeppli,  being two-parameter 
CPD's, 
differently from PCLD, satisfy all eq.~(16), with the corresponding 
$A_n$ coefficients shown in Table 1. 
Accordingly, for hierarchical models one has
$$
P_0(\Dy) = \exp \left[ \sum_{n=1}^{\infty} A_n
{[- \bar n(\Dy)]^n \over n!}  [\kappa_2(\Dy)]^{n-1}
\right] 
\eqno(17)
$$
with 
$$
\kappa_2(\Dy) = \int_{\Dy} dy_1  \int_{\Dy} 
dy_2 c_2(y_1,y_2) 
\eqno(18) 
$$
As already pointed out in the Introduction, 
these results represent the counterpart in terms of the average number 
of g-clans of the properties of the void scaling function 
${\cal V}(\Dy)$ discussed in \ref{1,14,15} 
in order to test normalized factorial cumulants hierarchical structure. 
Apparently the probability to detect no particles for a CPD in a given 
rapidity interval is controlled by and controls both the average number 
of g-clans of the full MD and the normalized factorial cumulant generating function in 
the same interval.  

\noindent
These results have a possible explanation in the existing connection 
between the $n$-particle and zero-particle probabilities 
as given by the following equation 
$$
P_n(\Dy) = {[- \bar n(\Dy)]^n \over n!} {\partial^n P_0(\Dy) \over 
\partial \bar n(\Dy)^n} 
\eqno(19)
$$
which can be obtained 
by allowing only $\bar n(\Dy)$ to vary in $P_0(\Dy)$ (all the other 
parameters of $P_0(\Dy)$ are taken fixed with respect to the 
$\bar n(\Dy)$ variation).  
It is to be noticed that from eq.~(19) one can deduce the following 
differential equation 
$$
(n+1)P_{n+1}(\Dy) - nP_n(\Dy) = - \bar n(\Dy) 
{\partial P_n(\Dy) \over \partial \bar n(\Dy)} 
\eqno(20)
$$
or, in terms of the corresponding 
generating function, $f(z;\Dy)$, 
$$
\bar n(\Dy) {\partial f(z;\Dy) 
\over \partial \bar n(\Dy)} = (z-1) {\partial f(z;\Dy) 
\over \partial z} 
\eqno(21)
$$
{\it i.e.}, $f(z;\Dy)$ depends on $z$ and $\bar n(\Dy)$ through the product
$\bar n(\Dy) (z-1)$ only.

From eq.~(12) 
all distributions  whose normalized
factorial cumulants do not depend on the average multiplicity, 
like for instance NB MD,  satisfy the above property (19).
It is interesting to remark that eq.~(19) is fulfilled by many 
distributions used in literature, like Poisson, NB, P\' olya-Aeppli, 
Thomas,  and 
all distributions which can be written 
as a positive weight superposition of Poisson 
distributions\ref{16} (Poisson transforms of a continuous distribution). 

The fact that eq~(11)  
 holds for any MD belonging to the class of CPD's, 
including NB MD, is of particular relevance. The importance of the result 
is enhanced by remembering the definition of void scaling function 
${\cal V}(\Dy)$ (see \ref{1}), which is just the inverse
of the average number of particles per g-clan, {\it i.e.}, 
$$
{\cal V}(\Dy) = {\Ngc (\Dy) \over 
{\bar n}(\Dy)} = {1 \over 
\ngc(\Dy)}  
\eqno(22)
$$
For the NB MD notice that $\Ngc (\Dy)$ and 
$\ngc(\Dy)$ of eq.s~(11) and (22) coincide with
eq. (8) and (9). 

Altogether above mentioned formulae 
show how deep and intriguing is the meaning of
what was believed for long time just a new parametrization of MD's for 
interpreting NB regularity;
generalized clan structure 
analysis turns out to be in general the analysis of voids or gaps 
properties in phase space and of the $n$-particle  
correlation function structure of the corresponding MD's.

\beginsection II. Rapidity gap probability from experimental 
multiplicity distributions 

It has been shown by Cugnon and Harouna\ref{17} that goodness of 
fits to 
experimental data on final charged particle MD's in terms of CPD's does 
not change much for different choices of the generating function $g(z)$. 
More precisely, all these fits are comparable with a NB fit. This fact 
can be interpreted as an indication that what matters more in the above
mentioned context is the CPD 
nature of the process (a two steps process) than the detailed structure 
of the MD obtained by fixing the generating function 
$g(z)$. In view of this remark, 
limiting the discussion just to one MD does not restrict the domain of 
validity of our general conclusions: they refer in an approximate sense  
-- with the appropriate warnings -- to the whole class of CPD's. 
The most natural choice is to discuss the NB MD. It is in fact quite clear 
that when one wants to apply results on CPD to the 
real world one meets necessarily just this distribution, whose 
approximate validity for describing final particles  MD's is 
well established. The interest of the application of results of 
Section I to the NB MD lies in the fact that our results should be 
thought common, in view of previous remark, to the whole class of CPD's. 

Deviations from NB behavior in symmetric rapidity intervals were indeed 
observed in $\bar p p$ reactions at c.m. energy \roots{900} and 
at c.m. energy \roots{1800} as in \ee\ annihilation at LEP 
(c.m. energy \roots{91}). A typical shoulder structure is seen in all 
these experiments. Shoulder effect is understood\ref{18} 
in \ee\ annihilation as 
the superposition of MD's of events of different topologies and each 
topology satisfies well NB behavior (see \ref{19} for a 
critical discussion of this point). 
In $\bar p p$ it has been 
proposed\ref{20} to describe the effect  by the superposition of 
two NB MD's, which is justified by the onset of a semi-hard component. 
The success of the fit is of course 
weakened in this case 
by the large number of parameters introduced. Notice that 
the analysis of the topology of the events is here not possible 
since UA5 Collaboration cannot measure particles' momenta. 
From our point of view, it is to be remarked 
that a linear superposition of two or
more CPD's allow to study the probability of detecting no particles in a 
given rapidity interval in terms of generalized clan structure 
analysis in the same interval. 
In fact eq.~(5) can be generalized as follows 
$$
f(z;\Dy) = \alpha f^{(1)}(z;\Dy) + (1-\alpha) f^{(2)}(z;\Dy) 
\eqno(23)
$$
with $\alpha$ a parameter controlling the relative weight of one 
distribution to the other. 
Accordingly, from eq.~(10), one obtains
$$
P_0(\Dy) = \alpha e^{-\Ngc^{(1)}(\Dy)} 
+ (1-\alpha) e^{-\Ngc^{(2)}(\Dy)}
\eqno(24)
$$
Consequently, observed deviations from NB behavior can be interpreted again 
in the same framework of CPD's. 
Shoulder effect in rapidity intervals could be analyzed in such 
terms also at \TEVATRON.

Let us now examine $P_0(\Dy)$ properties in the domain of validity 
of NB regularity. 
This analysis can be considered an example of the result discussed 
previously for CPD's as well as an alternative procedure for studying 
rapidity gap probabilities as they appear in the real world. 

In Figure 1 the rapidity gap probability in symmetric rapidity interval 
$\Dy$, $P_0(\Dy)$, obtained by performing a NB 
fit on the MD, is shown as a function of the width of the rapidity 
interval $\Dy$ for $hh$ collisions at different c.m. energies 
ranging from \roots{22}\ref{21} to \roots{546}\ref{4}. 
The rapidity gap probability decreases almost linearly for small rapidity 
intervals and then levels for larger rapidity intervals; this last behavior 
corresponds to the well-known bending of the average number  of clans 
observed experimentally at the border of phase space. This result has
been interpreted in \ref{6} as the effect of 
conservation laws, which become important at the boundary of phase space; 
this picture is supported by the experimental behavior of the 
average number of particles per clan, which after a quick increase 
reaches a maximum and then decreases in large rapidity intervals. 
It is to be pointed out that the rapidity gap probability in 
$hh$ collisions is approximately energy independent from \roots{22} up
to \roots{546}, as of course it is to be expected from the corresponding 
behavior in clan structure analysis.

\noindent 
A preliminary analysis of minimum bias events at \TEVATRON\  
(at c.m. energy \roots{1800}) has been 
actually performed by CDF Collaboration\ref{22}: it has been found 
that at this c.m. energy the shoulder structure becomes important not 
only in large rapidity intervals as it was found at c.m. energy 
\roots{900}\ref{23}, but 
also in smaller rapidity intervals. It could be interesting to 
investigate if this structure could be still explained in terms of the 
superposition of two different MD's, each of them of CPD type (for 
instance two NB MD's). In this case, as previously discussed, 
one would still be able to exploit CPD
structure to determine $P_0(\Dy)$ properties from the full MD and 
to compare this behavior with that shown in Figure 1.

\noindent 
A similar trend for the rapidity gap probability has been observed 
by D$\emptyset$ Collaboration\ref{24} 
for a different sample of events, {\it i.e.}, for dijet 
events with transverse energy of each jet greater than 30 GeV. 
This qualitative common structure is very remarkable since in this second 
case the selected process is a hard one. 

\noindent 
Notice that the use of CPD properties for studying the rapidity gap 
probability in minimum bias events at \TEVATRON\ would be really helpful 
for a deeper understanding of D$\emptyset$ results; in fact, this analysis could 
provide the estimate of the rapidity gap probability expected from the 
ordinary soft gluon radiation, {\it i.e.}, the estimate of the soft 
background contribution which is not under control so far in 
D$\emptyset$ data and can 
mask the detection of a hard production mechanisms, like, for instance, 
 hard pomeron exchange\ref{25}.

\noindent 
In view of the qualitative agreement between minimum bias 
results obtained in the framework of 
CPD properties (and in particular NB MD) and D$\emptyset$  
results for dijet events, it could be natural to assume that the 
resulting MD is of CPD type also for dijet events. 
Accordingly, one can determine $P_0(\Dy)$ in dijet 
events not only by looking directly at regions of phase space without 
any particle, but also, with a completely independent method, by applying
clan structure analysis to the full MD. Notice that this second
method should decrease the
statistical error on the rapidity gap probability, since it is based on the analysis
of the full sample of dijet events.

The structure of rapidity gap probability similar to that observed in 
$hh$ collisions is seen also in deep inelastic scattering\ref{26} 
(see Figure 2). 
It is interesting to remind\ref{6} that in 
this case the average number of clans has the same behavior as in 
$hh$ collisions, but the average number of particles per clan has a behavior 
similar to \ee\ annihilation, {\it i.e.}, clans in deep inelastic 
scattering are much smaller than in $hh$ collisions. 
This property could be relevant for the interpretation of the recent
experimental result found at \HERA\ref{27}, 
where an excess of events with a 
large rapidity gap and small multiplicity has been observed. 

Particular attention should be paid to \ee\ annihilation where, as well 
known, the average number of clans has a different slope with respect to 
the behavior shown in Figure 1 and 2. Therefore, one should expect a 
steeper slope for $P_0(\Dy)$. This is shown in Figure 3. 
This analysis is here limited to two-jet 
events in order to make possible the comparison between HRS\ref{28} and 
\delphi\ref{18} data. 
The study of rapidity gap probability 
is here simplified with respect to reactions with hadrons in the initial 
state, because in \ee\ annihilation a gap cannot be filled by 
particles produced by the fragmentation of the initial state remnants; 
this reaction constitutes therefore a ``clean'' 
environment to study the physics of rapidity gaps. 
In fact the information contained in Figure 3 can be used to estimate 
the two-jets events contribution to the background in the 
framework of the search proposed in \ref{29}.

Above results can be interpreted according to Theorems 2 and 3 discussed in
Section 1 also in terms
of correlations properties.
The effect of correlations can be tested indeed 
by looking at the difference between
the observed rapidity gap probability $P_0(\Dy)$, {\it i.e.}, 
eq.~(11) with $\bar N(\Dy)$ given by eq.~(8) as requested by NB behavior, 
and the rapidity gap probability corresponding to 
independent particle production, $e^{-\bar n(\Dy)}$, 
{\it i.e.}, to a Poissonian 
distribution with the same average number of charged particles 
$\bar n(\Dy)$, to which $\bar N(\Dy)$ reduces in this case. 

\noindent 
In Table 2 experimental results for $e^{- \bar N(\Dy)}$ and
$e^{- \bar n(\Dy)}$ for different reactions in two 
fixed rapidity intervals are shown:
the effect of correlations is remarkable in $hh$ collisions where the rapidity gap
probability is larger by many order of magnitude with respect to 
the hypothetical Poissonian behavior; the difference is narrower in \ee\
annihilation two-jets events, 
confirming the quasi-Poissonian behavior of MD's.
Notice that 
the direct comparison between $hh$ collisions and \ee\ annihilation 
two-jets events shows that the rapidity
gap probability is larger in $hh$ collisions than in \ee\ annihilation.
This result agrees with previous interpretation 
of the parameter $1/k(\Dy)$ of NB MD as an aggregation parameter\ref{6}:
$$
{1 \over k} = {{\cal P}(n=2,N=1) \over {\cal P}(n=2,N=2)} 
\eqno(25)
$$
(${\cal P}(n,N)$ is the probability to produce $n$ particles 
distributed in $N$ clans), as well as with the relation of $1/k(\Dy)
$ with two-particle correlation function (see eq.~(18)).
In fact, suppose we compare two different reactions having the same 
average number of particles in a given rapidity interval $\Dy$; it 
is clear that to larger rapidity gap probability corresponds at fixed 
number of particles more aggregation among final particles, 
{\it i.e.}, larger values of $1/k(\Dy)$ as can be seen just by 
inspection of eq.s~(11) and (8). At the same time larger 
aggregation corresponds to larger two-particle correlations as can be 
noticed again just by inspection of eq.~(18). 

\beginsection Conclusions

The parameters introduced some time ago by L\'eon Van Hove and one of 
the present authors in order to interpret the wide occurrence in all 
classes of reactions of NB regularity have been found to possess a deep 
and intriguing physical meaning; they are related to $n$-particle 
correlation functions and can be used to test their eventual 
hierarchical structure, as it has been already anticipated in part in our 
previous work on void scaling function. 
This finding is shown to be common to all 
classes of CPD's, where the concept of generalized average number of 
clans, $\Ngc$, and of average number of 
particles per g-clan, $\ngc$, can be defined. 
This fact is of particular relevance: it points out the two steps nature 
of the physical process under investigation, which seems to be not much 
influenced by the detailed structure of the MD of particles  
inside an average g-clan, {\it i.e.}, by the structure of 
the second step in the process. 

The paper should be considered 
a contribution to the integrated description of 
$n$-particles  correlation function and MD's in 
rapidity intervals. The link is represented by the probability to detect 
no particles in different rapidity regions of phase space. 
Accordingly an alternative approach is proposed in order to determine 
rapidity gap probability in terms of the average number of clans in the 
rapidity interval considered.

The detailed study of rapidity gap probability 
in the domain of validity of NB regularity 
reveals interesting features. In particular, one should mention the 
energy independence of rapidity gap probability for each class of 
reactions and its leveling for large rapidity intervals in $hh$ and $lh$ 
collisions. 
These remarks can be useful and inspiring in order to discuss data on 
rapidity gap probability at \TEVATRON\ and \HERA, where usually a  
sample of events different from minimum bias is selected. 

\beginsection References

\item{[1]}
S. Lupia, A. Giovannini, R. Ugoccioni,
Z. Phys.\ C59 (1993) 427

\item{[2]}
G.J.H. Burgers, C. Fuglesang, R. Hagedorn, V. Kuvshinov, 
Z. Phys. C46  (1990) 465 

\item{[3]}
A.\ Giovannini, Il Nuovo Cimento 10A (1972) 713;

\item{[4]}
G.J.\ Alner et al., UA5 Collaboration, Phys.\ Lett.\ B160 (1985) 193

\item{[5]}
N.\ Schmitz, in Multiparticle Dynamics (Festschrift for L\'eon Van Hove),
La Thuile, Italy, eds. A.\ Giovannini and W.\ Kittel,
World Scientific, Singapore 1990, p.~25

\item{[6]} 
L.\ Van Hove, A.\ Giovannini, in Proceedings of XVII International
Symposium on Multiparticle Dynamics, Seewinkel, eds. M.\ Markitan et al.,
World Scientific, Singapore 1987, p.~561; 
\hfill\break 
A. Giovannini, L. Van Hove, Z.\ Phys.\ C30 (1986) 391

\item{[7]}
N.L. Johnson, S. Kotz, A.W. Kemp, ``Univariate discrete distributions'', Second
Edition, J. Wiley \& Sons, N.Y., 1993

\item{[8]}
A. Capella, A.V. Ramallo, Phys. Rev. D37 (1988) 1763 

\item{[9]}
M. Biyajima , T. Kawabe, N. Suzuki, Phys. Lett. B189 (1987) 466

\item{[10]} 
P. Carruthers, C.C. Shih, Phys. Lett. B128 (1983) 242; 
\hfill\break
M. Biyajima, Progr. Theor. Phys. 69 (1983) 966

\item{[11]}
S. Hegyi, Phys. Lett. B318 (1993) 642

\item{[12]}
S. Lupia, A. Giovannini, R. Ugoccioni, 
in Proceedings of the Conference ``Hadron Structure 93''
(Bansk\'a \v{S}tiavnica, Slovakia, 1993), eds. S.~Dubni\v{c}ka and 
A.Z.~Dubni\v{c}kov\'a, Slovak Academy of Sciences, Bratislava, 1993,
p.~198

\item{[13]}
P. Carruthers, I. Sarcevic, Phys. Rev. Lett. 63 (1987) 1562; 
\hfill\break 
L. Van Hove, Phys.\ Lett.\ B242 (1990) 485

\item{[14]}
S. Hegyi, Phys.\ Lett.\ B271 (1992) 214

\item{[15]}
E.A.\ De Wolf, in: Fluctuations and Fractal Structure, 
Proceedings of the Ringberg Workshop on Multiparticle
Production (Ringberg Castle, Germany, 1991), eds.~R.C.~Hwa, 
W.~Ochs and N.~Schmitz, World Scientific,
Singapore, 1992, p.~222

\item{[16]} 
P. Carruthers, C.C. Shih, Int. J. Mod. Phys. A2 (1987) 1447 

\item{[17]}
J. Cugnon, O. Harouna, Europhys. Lett. 4 (1987) 1127 

\item{[18]}
P.\ Abreu et al., \delphi\ Collaboration, Z.\ Phys.\ C56 (1992) 63

\item{[19]}
F. Bianchi, A. Giovannini, S. Lupia, R. Ugoccioni,
Z. Phys.\ C58 (1993) 71

\item{[20]}
C.\ Fuglesang, in Multiparticle Dynamics (Festschrift for L\'eon
Van Hove), La Thuile, Italy, eds. A.\ Giovannini and W.\ Kittel,
World Scientific, Singapore 1990, p.~193

\item{[21]}
A. Breakstone et al., Il Nuovo Cimento 102A (1989) 1199

\item{[22]}
F. Rimondi, preprint FERMILAB-Conf-93-359-E, November 1993, 
to be published in the Proceedings of the XXIII International Symposium
on Multiparticle Dynamics, Aspen, CO, USA, 12-17 September 1993

\item{[23]}
R.E.\ Ansorge et al., UA5 Collaboration, Z.\ Phys.\ C43 (1989) 357

\item{[24]}
S. Abachi et al., D0 Coll., Phys. Rev. Lett. D72 (1994) 2332

\item{[25]}
J.D. Bjorken, Phys. Rev. D47  (1993) 101 

\item{[26]}
M.\ Arneodo et al., EMC Collaboration, Z.\ Phys.\ C35 (1987) 335

\item{[27]}
M. Derrick et al., ZEUS Coll., Phys. Lett. B315 (1993) 481;
\hfill\break
M. Derrick et al., ZEUS Coll., DESY preprint 94-063, April 1994; 
\hfill\break
A. De Roeck, H1 Collaboration, Proc. Int. Europhysics Conf. on HEP, 
Marseille 1993, Eds. J. Carr and M. Perrottet, p.~791.

\item{[28]}
M.\ Derrick et al., HRS Collaboration, Phys.\ Rev.\ D34 (1986) 3304

\item{[29]}
J.D. Bjorken, S.J. Brodski, H.J. Lu, Phys. Lett. B286 (1992) 153; 
\hfill\break 
H.J. Lu, preprint U. Maryland,  U.MD/94-073

\vfill\eject

\beginsection Figure Captions

\noindent 
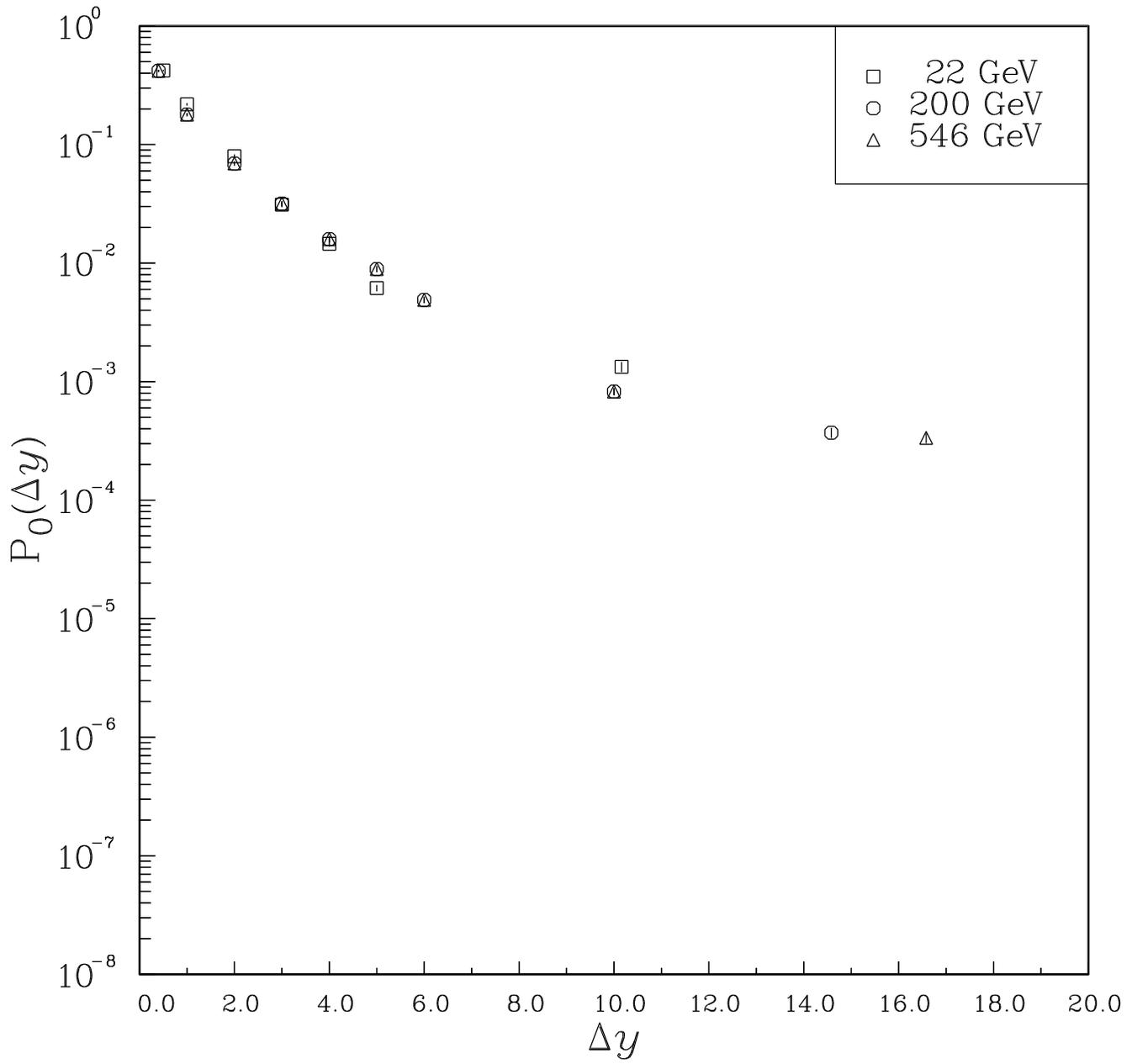
{\bf Fig. 1}: 
Rapidity gap probability $P_0(\Dy)$ as a function of the rapidity width 
$\Dy$ obtained from NB fits in $hh$ collisions at different c.m. 
energies \roots{22}, \roots{200} and \roots{546}.

{\bf Fig. 2}: 
Rapidity gap probability $P_0(\Dy)$ as a function of the rapidity width 
$\Dy$ obtained from NB fits in deep inelastic scattering 
for various intervals of the total hadronic energy $W$ as indicated in 
the Figure.

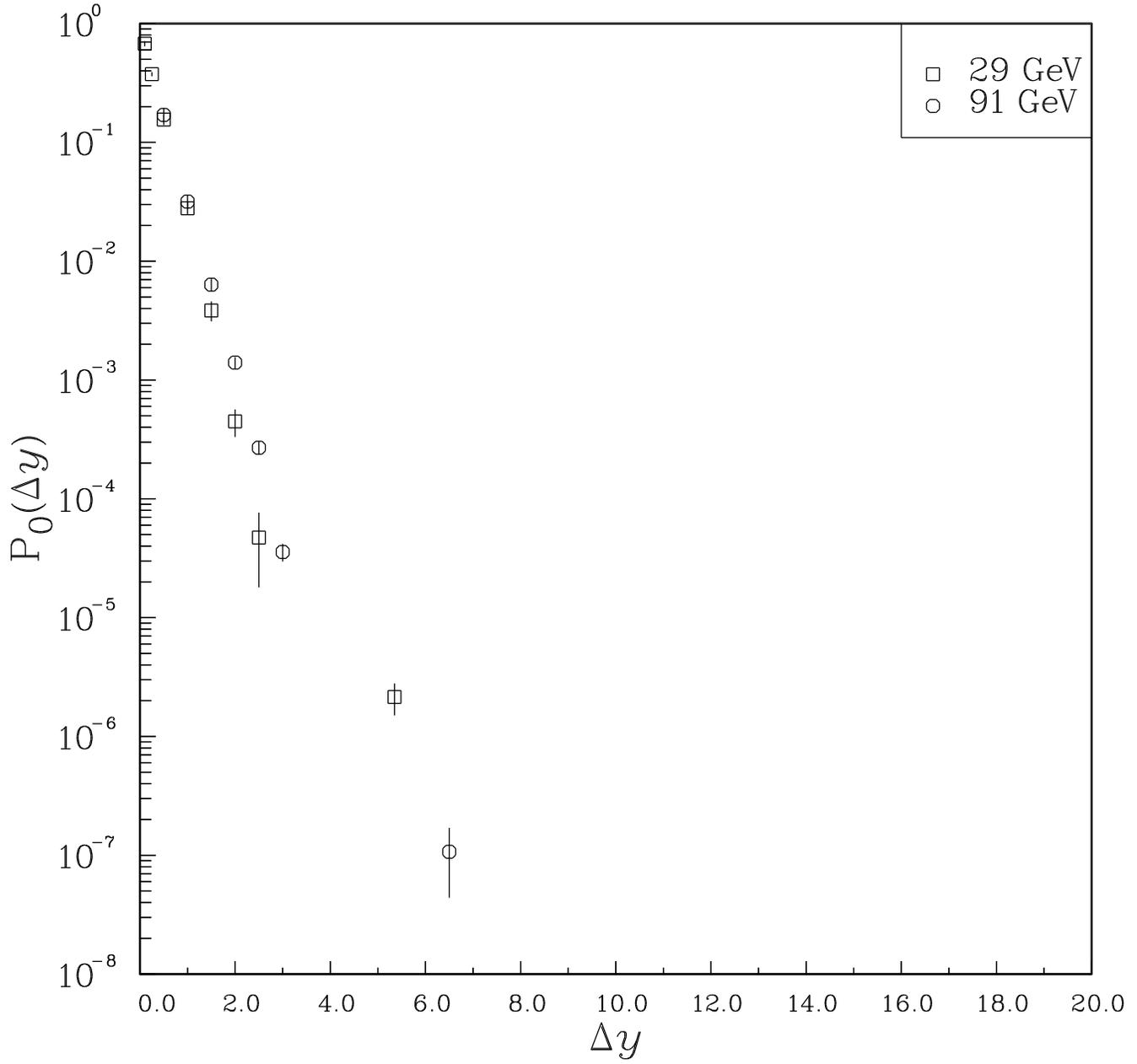
{\bf Fig. 3}: 
Rapidity gap probability $P_0(\Dy)$ as a function of the rapidity width 
$\Dy$ obtained from NB fits to the sample of two-jets events 
in \ee\ annihilation at c.m. 
energies \roots{29} and \roots{91}.

\beginsection Table Captions 

{\bf Tab. 1}: generating function $f(z)$ of final particles MD, 
average number of g-clans, $\Ngc$, and generating function of particles 
MD inside an average g-clan, $g(z)$,   
for NB, Thomas, P\' olya-Aeppli distributions as a function of  
the average number of particles $\bar n$ and  the 
second-order normalized factorial cumulant $\kappa_2$ of each distribution.  
In the last column $A_n$ coefficients (see eq.~(16)) of the above 
mentioned distributions are indicated. 
Generating function of the MD, average number of g-clans and 
generating function of the MD inside a g-clan for the PCLD are also 
shown in terms of its three parameters $A$, $B$, $C$. It should be added 
that eq.~(16) is not valid in this case.

{\bf Tab. 2}: Comparison in two 
different rapidity intervals of rapidity gap probability obtained via the 
average number of clans from a NB fit, $\exp (-\bar N)$,  
and the rapidity gap  probability expected for a Poissonian distribution 
of the same average number of particles, $\exp (-\bar n)$,  
for the reactions indicated in the Table. 
Values of the average number of clans $\bar N$ and of the 
average multiplicity $\bar n$ in the same intervals for the same 
reactions are also shown.

\bye